\documentclass[%
 reprint,
superscriptaddress,
 amsmath,amssymb,
 aps,
]{revtex4-2}

\usepackage{graphicx}%
\usepackage{dcolumn}%
\usepackage{bm}%

\usepackage{pifont}%

\begin{document}

\preprint{APS/123-QED}

\title{Generalised All-Optical Cat Correction}

\author{Ari John Boon}
\affiliation{Department of Engineering Physics, École polytechnique de Montréal, Montréal, QC, H3T 1J4, Canada}

\author{Olivier Landon-Cardinal}
\affiliation{École de technologie supérieure, Université du Québec, Montréal, QC, H3C 1K3, Canada}

\author{Nicol{\'a}s Quesada}
\affiliation{Department of Engineering Physics, École polytechnique de Montréal, Montréal, QC, H3T 1J4, Canada}

\date{\today}

\begin{abstract}

We have generalised an all-optical telecorrection protocol for the higher orders of the cat code, and show that with these higher orders we can achieve target performance at substantially reduced iteration counts at the cost of a higher mean photon-number.
We also introduce a probabilistic scheme for correcting deformation of the state, which highlights two interesting abilities of telecorrection: to encode new sets of transformations, and to change the basis of the code.
We find that for a target channel fidelity of $99.9\%$ over a channel with $1\text{ dB}$ of loss, a third-order cat code requires $70$ times fewer telecorrection iterations than a first-order one, at a cost of a $3.6$-fold increase in mean photon-number.
\end{abstract}

\maketitle

\section{Introduction}

Photonics is the natural platform for quantum communications, but is subject predominantly to the effects of loss, which will quickly obliterate the quantum state one wishes to transmit.
As a bosonic system, optical quantum communications can benefit from the numerous bosonic codes that have been shown to suppress the effects of loss \cite{gottesman64encoding, PhysRevA.59.2631, michael2016new, bergmann2016quantum, albert2018performance, grimsmo2020quantum, totey2023performance, rmlm-vbfd}.
Among these codes, however, their performance against loss has frequently been assessed without consideration for practical photonic implementations, where the deterministic operations and measurements one has access to are considerably limited.
The cat code is in the class of rotation-symmetric bosonic (RSB) codes, which has a clear mechanism for loss-correction and an explicit implementation, but relies on access to two-mode rotations \cite{grimsmo2020quantum}.
Optically, the only easily accessible two-mode operation is the beamsplitter, while deterministic single-mode operations are limited to Gaussian operations.
The most accessible optical nonlinearity comes from photon-number measurement, which demands cryogenic cooling for what is otherwise a room-temperature system.

More recent works have begun to consider optical implementation more explicitly,
as all-optical loss-correction circuits have been proposed for the GKP \cite{marqversen2025performance} and cat codes \cite{hastrup2022all, su2022universal}.
For cat codes in particular, not only are they well-behaved optically on account of being constructed from coherent states, but it has been shown that by iteratively correcting them one can achieve seemingly arbitrary performance with a relatively simple optical circuit \cite{hastrup2022all}.
However, in that proposal, the cost of arbitrary performance is a steep increase in the number of iterations and an increasing mean photon-number.
This means that even with its simple three-mode circuit, minimising error rates demands an exceedingly large number of photon-number measurements and ancilla states over the course of transmission.

The optical schemes proposed in Refs.~\cite{hastrup2022all} and \cite{su2022universal} were limited to cat codes which can only be explicitly corrected from single-photon losses.
These represent only the lowest-order of loss-correcting cats \cite{bergmann2016quantum}, which have been shown to be inferior in terms of absolute performance versus loss relative to their higher-order counterparts \cite{albert2018performance, totey2023performance}.
As higher-order cats retain their relatively compact and optically-favourable basis of coherent states, it is natural to assume that they too have a straightforward way of being implemented optically.
Here we show that not only is this possible, but it is advantageous, as higher-order cat codes can be used to greatly reduce the number of circuit iterations one requires, albeit at the cost of working in a higher mean photon-number regime.
As research into realising high-resolution photon-number measurements \cite{eaton2023resolution, dalbec2025accurate, li2025boosting} and photonic state-generation \cite{fiuravsek2005conditional, Thekkadath2020engineering, yao2024riemannian, crescimanna2024seeding, takase2024generation, forbes2025heralded} is ongoing and fruitful, our results show a valuable new axis along which to distribute the resource cost of telecorrection.

The outline of our paper is as follows.
In Sec.~\ref{sec:cats}, we describe the structure and properties of the cat code, in particular the origin of their loss capacity and sources of inherent error.
In Sec.~\ref{sec:syndromes}, we provide the circuit used for cat telecorrection, and write the generalised syndrome which indicates logical $X$ and $Z$ errors and quantifies additional errors in the state.
In Sec.~\ref{sec:performance}, we numerically compare the performance of different orders of the cat code in terms of the mean channel fidelity, and find the parameter requirements for each code order to achieve certain target fidelities.
In Sec.~\ref{sec:dedeformation}, we propose a scheme to probabilistically correct additional errors in the cat states that cannot be removed deterministically, for which we outline two interesting properties of the telecorrection scheme in Sec.~\ref{sec:ancilla_tricks}: the ability to encode new transformations, and to change the basis of our input state.
In Sec.~\ref{sec:conclusion}, we summarise our results, highlighting the experimental necessities for realising higher-order cat correction based on our generalised protocol.

\section{Multicomponent cat codes}
\label{sec:cats}

The logical states of the cat code are constructed from superpositions of coherent states, 
\begin{equation}
\vert e^{i \phi} \alpha \rangle
=
e^{-\alpha^2/2}
\sum^{\infty}_{n=0}
\frac{(e^{i\phi}\alpha)^n}{\sqrt{n!}}
\vert n \rangle,
\label{eqn:coherent}
\end{equation}
for $\alpha \in \mathbb{R}_{>0}$, where $\vert n \rangle$ is the Fock state for photon-number $n$.
Coherent states are the eigenstates of the annihilation operator,
\begin{equation}
\hat{a} \vert e^{i\phi} \alpha \rangle
=
e^{i\phi} \alpha \vert e^{i\phi} \alpha \rangle,
\label{eqn:aa}
\end{equation}
leading the expectation value of the photon number to be
\begin{equation}
\langle e^{i\phi} \alpha \vert \hat{n} \vert e^{i\phi} \alpha \rangle = \alpha^2,
\end{equation}
where $\hat{n} = \hat{a}^{\dagger} \hat{a}$.
The number of coherent states in a cat state determines the degree of discrete loss events that can be recovered from \cite{bergmann2016quantum}.
For a cat code with $L$-photon loss capacity, we define the logical states as
\begin{align}
\vert \tilde{0}_L \rangle
&=
\frac{1}{\sqrt{\mathcal{N}_0}}
\sum^{2 L + 1}_{m=0}
\vert e^{m \pi i / (L+1) } \alpha \rangle,
\label{eqn:codewords_coherent0}
\\
\vert \tilde{1}_L \rangle
&=
\frac{1}{\sqrt{\mathcal{N}_1}}
\sum^{2 L + 1}_{m=0}
(-1)^m
\vert e^{m \pi i / (L+1) } \alpha \rangle
\label{eqn:codewords_coherent1}
\end{align}
with a tilde indicating the logical basis, and $L$ its order, though we will leave these implicit when possible.
The loss capacity of the cat code arises from the Fock distance between the codewords, which can be re-written as
\begin{align}
\vert \tilde{0}_L \rangle
&\propto
\sum^{\infty}_{n=0}
\frac{\alpha^{2 n (L + 1)}}{\sqrt{[2 n (L + 1)]!}}
\vert 2 n (L + 1) \rangle,
\\
\vert \tilde{1}_L \rangle
&\propto
\sum^{\infty}_{n=0}
\frac{\alpha^{(2 n + 1) (L + 1)}}{\sqrt{[(2 n + 1) (L + 1)]!}}
\vert (2 n + 1) (L + 1) \rangle,
\label{eqn:codewords_fock}
\end{align}
where we see that our codewords are separated in Fock space by a distance $(L+1)$, and that at a given order the qubit will be distinguished by support on every $n \equiv 0 \bmod (L+1) $.
Therefore photon losses up to $L$ can be identified by the Fock support of the lossy qubit.
In the $L=1$ code, for instance, there is a tolerance for up to a single-photon loss, with even Fock support corresponding to the lossless state, and odd Fock support corresponding to the state after the single-photon loss.

\begin{figure}[t]
\includegraphics{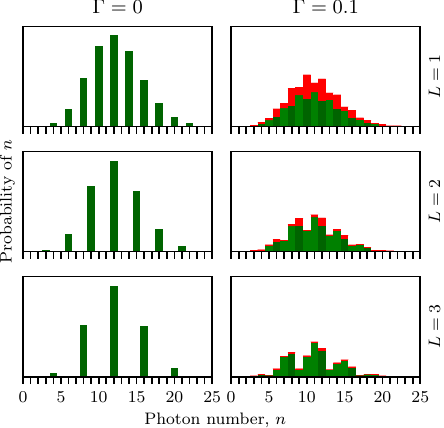}
\caption{\label{fig:fock_overlap} Fock probabilities for lossless (left) and lossy (right) cat states at different code orders, with cumulative contributions from loss orders beyond the loss capacity of each code ($l>L$) shown in red, taken here to a cutoff in the loss channel of $l=10$.
The red therefore corresponds to the magnitude of uncorrectable loss error, where we see that at higher $L$ these errors are much less significant.
All states are prepared at amplitudes of $\alpha=3.5$.}
\end{figure}

However, unlike the loss capacity of the cat code, the Kraus decomposition of the loss channel is not finite, where for a given magnitude of energy loss, $\Gamma \in [0,1]$, we have
\begin{equation}
\mathcal{L}_{\Gamma} (\rho)
=
\sum^{\infty}_{l=0}
\hat{B}^{(l)}_{\Gamma}
\rho
\left[
\hat{B}^{(l)}_{\Gamma}
\right]^{\dagger},
\label{eqn:loss_channel}
\end{equation}
whose Kraus operators are
\begin{equation}
\hat{B}^{(l)}_{\Gamma}
=
\sqrt{\frac{\Gamma}{1-\Gamma}}^{\,l}
\,
\frac{\hat{a}^l}{\sqrt{l!}}
\,
\sqrt{1-\Gamma}^{\,\hat{n}}.
\label{eqn:loss_kraus}
\end{equation}
Orders $l > L$ of Eq.~(\ref{eqn:loss_channel}) will produce uncorrectable errors in an $L$-loss cat code, as their Fock support is degenerate with the correctable orders of loss.
One must therefore work in a regime where total loss is low enough for errors to be negligible at orders beyond the loss capacity of the code.
Fig.~\ref{fig:fock_overlap} shows that for a fixed amount of loss and equivalent state amplitudes, successive orders of the cat code see significantly reduced contributions associated with beyond-capacity loss orders.

\begin{figure}[t]
\includegraphics{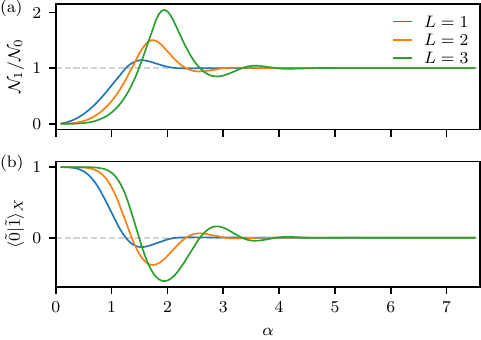}
\includegraphics{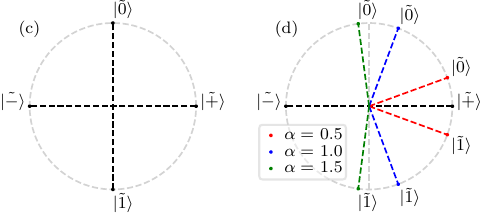}
\caption{\label{fig:norm_overlap_mismatch} The ratio between the codeword normalisation constants in our choice of basis (a), and the codeword overlap in the basis of Ref.~\cite{hastrup2022all} (b), shown for various code orders.
These quantities are related to inherent forms of error in either basis, and are directly related by Eq.~(\ref{eqn:norm_overlap_mismatch}).
At the bottom we show the $L=1$ codewords on a slice of the Bloch sphere in our choice of basis (c) and that of Ref.~\cite{hastrup2022all} or Eq.~(\ref{eqn:Rx_cats}) (d), where in (d) the codewords $\vert \tilde{0} \rangle$ and $\vert \tilde{1} \rangle$ are shown to be non-orthogonal.}
\end{figure}

However this does not make higher-order cat codes a universally better choice, as the higher orders suffer more acutely from errors inherent to the correction of the code, independent of loss, which we will elaborate on in Sec.~\ref{sec:syndromes}.
These correction errors take the form of a deformation related to the mis-match of the codeword normalisation constants, $\mathcal{N}_0$ and $\mathcal{N}_1$, which is generally worse the lower the state amplitude \cite{albert2018performance,bergmann2016quantum}, while errors due to loss are worse with increasing state amplitude due to the Fock-damping operator $\sqrt{1-\Gamma}^{\hat{n}}$ in Eq.~(\ref{eqn:loss_kraus}).
This leads to optimal performance at amplitudes where these two effects are balanced.
Normalisation mis-match is shown in Fig.~\ref{fig:norm_overlap_mismatch}(a) to both be more extreme and to persist to higher amplitudes in higher-order cat codes, which in turns pushes their optimal amplitudes higher, to be seen in Sec.~\ref{sec:performance}.

Notably, this is complementary to the problem of codeword overlap seen in the alternate basis used by Ref.~\cite{hastrup2022all}, whose codewords we specify in Appendix~\ref{app:bases}, Eq.~(\ref{eqn:Rx_cats}).
Codeword overlap arises in that basis due to the non-orthogonality of its codewords $\vert \tilde{0} \rangle$ and $\vert \tilde{1} \rangle$, shown in Fig.~\ref{fig:norm_overlap_mismatch}(c), and corresponds to a degree of indistinguishability for states defined that way.
This overlap in the alternate basis is plotted in Fig.~\ref{fig:norm_overlap_mismatch}(b) for increasing code order, where we see it follows similar trends as normalisation mis-match in our choice of basis.
In Appendix~\ref{app:bases} we derive for all orders a direct relationship between codeword normalisation mis-match in our choice of basis and codeword overlap in the basis of Ref.~\cite{hastrup2022all}, 
\begin{equation}
\frac{\mathcal{N}_0}{\mathcal{N}_1}
=
\frac{ 1 + \langle \tilde{0} \vert \tilde{1} \rangle_X }{ 1 - \langle \tilde{0} \vert \tilde{1} \rangle_X },
\label{eqn:norm_overlap_mismatch}
\end{equation}
where $\langle \tilde{0} \vert \tilde{1} \rangle_X$ is the codeword overlap for the basis of Ref.~\cite{hastrup2022all}.
With this we show that when the source of inherent error is small or large in one basis, it is necessarily also so in the other.
We also show numerically that for $L=1$ the differences in performance between these bases are relegated to the sub-optimal regime.
State indistinguishability can be easily quantified via codeword overlap, and we will see in Sef.~\ref{sec:syndromes} that the mechanism by which codeword normalisation mis-match introduces error is independent of code order.
Thus, quantifying both properties as in Eq.~(\ref{eqn:norm_overlap_mismatch}) and Fig.~\ref{fig:norm_overlap_mismatch} gives a reasonable notion of the severity of their associated errors.

\section{Generalised cat telecorrection}
\label{sec:correction}

Using the telecorrection protocol proposed in Ref.~\cite{hastrup2022all} for the $L=1$ cat code, we find that the same optical circuit can be used to correct higher orders of the cat code up to an adjustment in the syndromes and the preparation of a higher-order ancilla.
The protocol otherwise remains the same.

\begin{figure}[!ht]
\includegraphics{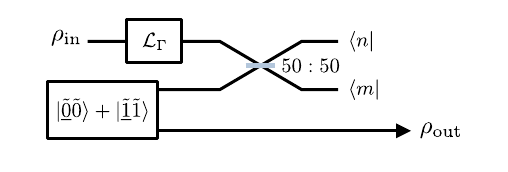}
\caption{\label{fig:protocol} The cat telecorrection protocol proposed in Ref.~\cite{hastrup2022all}.
Here an underbar is used to indicate that the first mode of the ancilla is prepared at a lower amplitude matching that of the input state after loss.}
\end{figure}

As illustrated in Fig.~\ref{fig:protocol}, the input mode is mixed via a real symmetric ${50}{:}{50}$ beamsplitter with the first mode of an ancilla
\begin{equation}
\vert \tilde{\Phi}^{+}_L \rangle
=
\frac{\vert \underline{\tilde{0}}_L \tilde{0}_{L} \rangle + \vert \underline{\tilde{1}}_L \tilde{1}_{L} \rangle}{\sqrt{2}},
\end{equation}
where the ancilla is composed of order-$L$ cat states, with its first mode prepared at amplitude $\sqrt{1 - \Gamma} \alpha$ to match that of the lossy input state, with the reduced amplitude indicated by an underbar.
Photon-number measurements, $(n,m)$, are performed on the first two modes---that of the input state and the reduced-amplitude mode of the ancilla---and act as the syndrome for the state teleported into the logical basis of the remaining ancilla mode.

Teleportation into the ancilla's remaining mode both restores it to our desired basis and corrects the amplitude damping.
Not only does this mechanism correct our state, but it is also convenient numerically, as it allows us to express for every syndrome $(n,m)$ a direct mapping between a logical input state and a logical output state.
This mapping is derived in Appendix~\ref{app:teleportation}, Eq.~(\ref{eqn:rho_out}).

\subsection{Error syndromes}
\label{sec:syndromes}

Due to our choice of basis relative to Ref.~\cite{hastrup2022all}, we see a relative exchange of syndromes for logical $X$ and $Z$ errors at $L=1$.
We also utilise a numerical mapping of the Pauli errors, as the relations for $Z$ syndromes are non-trivial at orders $L > 1$.

In Appendix~\ref{app:syndrome} we derive an expression for the $2~\times~2$ transformation associated with each order of the loss channel, $l$, and each pair of photon-number measurements, $(n,m)$.
Up to order $l=L$ in the channel, due to the Fock support of each loss event up to that order being unique, a single order of loss is associated with a given pair of photon-number measurements, $(n,m)$, following the rule
\begin{equation}
l \equiv L (n+m) \bmod (L+1).
\end{equation}
This allows us to truncate the channel and quantify the error associated with each syndrome, $(n,m)$, as follows:
\begin{widetext}
\begin{eqnarray}
S_{KJ} (n,m)
=
\sum^{2L+1}_{j,k=0}
&&e^{i j l \pi / (L+1)}
\,
\frac{\bigl( \sqrt{\Gamma} \alpha \bigr)^l}{\sqrt{l!}}
\,
\frac{(-1)^{jJ+kK}}{\sqrt{2 \mathcal{N}_J \underline{\mathcal{N}}_K}}
\,
\frac{e^{(\Gamma-2)\vert \alpha \vert^2 / 2}}{\sqrt{n!m!}}
\,
\left[
\sqrt{\frac{1 - \Gamma}{2}} \alpha
\right]^{n+m}
\nonumber
\\
&&\cdot
\left[ 
e^{i j \pi / (L+1)}
+
e^{i k \pi / (L+1)}
\right]^n
\,
\left[
e^{i k \pi / (L+1)}
-
e^{i j \pi / (L+1)}
\right]^m,
\label{eqn:syndrome}
\end{eqnarray}
\end{widetext}
where $S_{KJ}(n,m)$ is a scalar element of $S(n,m)$, which is a $2 \times 2$ transformation matrix defined on the indices of the logical basis, $J,K \in \{0,1\}$; the underbar in $\underline{\mathcal{N}}_K$ indicates that the normalisation constant is that of a cat state prepared at reduced amplitude---the first mode of the ancilla.
Syndromes calculated this way are only accurate up to a certain magnitude of loss for a given amplitude and order of cat code, but this is the same regime in which we wish to use the code: one in which orders of loss beyond the capacity of our code do not significantly contribute.

\begin{figure}[t]
\includegraphics{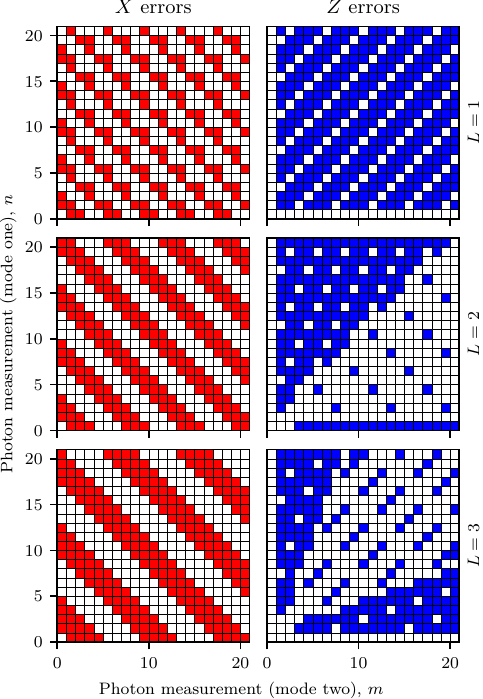}
\caption{\label{fig:pauli_maps} Maps of the logical $X$ (red) and $Z$ (blue) errors in terms of photon-number measurements, $(n,m)$, on either mode, determined numerically from Eq.~(\ref{eqn:syndrome}).
White indicates that the respective error is not present.
$X$ and $Z$ errors may occur simultaneously, and are always correctable.
Transmission failure and deformation can be mapped similarly.
}
\end{figure}

We observe three categories of $S(n,m)$: explicit transmission failure cases where the logical information is lost, deformed off-diagonal transformations, and deformed diagonal transformations; we show examples of these here, respectively, for $L=1$, $\alpha=2$, and $\Gamma=0$:
\begin{align}
S(0,0)
&=
\begin{pmatrix}
\frac{8 \sqrt{2}}{\sqrt{\mathcal{N}_0 \underline{\mathcal{N}}_0} e^4} & 0 \\
0 & 0 
\end{pmatrix},
\label{eqn:failure}
\\
S(2,0)
&=
\begin{pmatrix}
0 & \frac{16}{\sqrt{\mathcal{N}_1 \underline{\mathcal{N}}_0} e^4} \\
\frac{16}{\sqrt{\mathcal{N}_0 \underline{\mathcal{N}}_1} e^4} & 0 
\end{pmatrix}
=
D(2,0)
\,
\sigma_x,
\label{eqn:bit_flip}
\\
S(2,2)
&=
\begin{pmatrix}
\frac{32 \sqrt{2}}{\sqrt{\mathcal{N}_0 \underline{\mathcal{N}}_0} e^4} & 0 \\
0 & \frac{-32 \sqrt{2}}{\sqrt{\mathcal{N}_1 \underline{\mathcal{N}}_1} e^4} 
\end{pmatrix}
=
D(2,2)
\,
\sigma_z,
\label{eqn:phase_flip}
\end{align}
where the deformations are
\begin{align}
D(2,0)
&=
\begin{pmatrix}
\frac{16}{\sqrt{\mathcal{N}_0 \underline{\mathcal{N}}_1} e^4} & 0 \\
0 & \frac{16}{\sqrt{\mathcal{N}_1 \underline{\mathcal{N}}_0} e^4} 
\end{pmatrix},
\\
D(2,2)
&=
\begin{pmatrix}
\frac{32 \sqrt{2}}{\sqrt{\mathcal{N}_0 \underline{\mathcal{N}}_0} e^4} & 0 \\
0 & \frac{32 \sqrt{2}}{\sqrt{\mathcal{N}_1 \underline{\mathcal{N}}_1} e^4} 
\end{pmatrix},
\end{align}
which we can further reduce to deformation ratios,
\begin{equation}
d_{2,0}
=
\sqrt{\frac{\mathcal{N}_0 \underline{\mathcal{N}}_1}{\mathcal{N}_1 \underline{\mathcal{N}}_0}},
\quad
d_{2,2}
=
\sqrt{\frac{\mathcal{N}_0 \underline{\mathcal{N}}_0}{\mathcal{N}_1 \underline{\mathcal{N}}_1}}.
\end{equation}
Eq.~(\ref{eqn:failure}) corresponds to complete loss of the qubit, while in Eqs.~(\ref{eqn:bit_flip}) and (\ref{eqn:phase_flip}) we find that the transformations contain logical $X$ and $Z$ errors up to some additional deformation.
A mapping of logical $X$ and $Z$ errors extracted this way from Eq.~(\ref{eqn:syndrome}) is shown in Fig.~\ref{fig:pauli_maps}, and is used to determine the appropriate Pauli corrections to apply.
Pauli-correction does not correct deformation, which we see is determined by ratios of normalisation constants, $\mathcal{N}_J$ and $\underline{\mathcal{N}}_K$.

Deformation occurs here as a result of weak Pauli $Z$ measurements, similar to the weak Pauli $X$ measurements originally observed in Ref.~\cite{hastrup2022all}, but more prevalent here due to the normalisation constant mis-match.
However, because our basis does not suffer from non-orthogonality, this is not comparatively detrimental.
Deformation constitutes a violation of the Knill-Laflamme conditions \cite{knill1997theory},
\begin{equation}
\langle \tilde{J} \vert P \vert \tilde{J'} \rangle
= \epsilon_{P} \delta_{J,J'},
\quad
J,J' \in \{ 0, 1 \},
\label{eqn:KL_conditions}
\end{equation}
with $P$ some correctable error, and $\epsilon_P$ independent of $J$ and $J'$.
Although our choice of basis preserves orthogonality under a Pauli-corrected error, $S'(n,m)$,
\begin{equation}
\langle \tilde{0} \vert S'(n,m) \vert \tilde{1} \rangle
=
0,
\label{eqn:KL_orthog}
\end{equation}
where the appropriate $X$ or $Z$ operation has been applied, its remaining deformation has the following effect:
\begin{equation}
\langle \tilde{0} \vert S'(n,m) \vert \tilde{0} \rangle
\neq
\langle \tilde{1} \vert S'(n,m) \vert \tilde{1} \rangle,
\label{eqn:KL_deform}
\end{equation}
as weak Pauli measurement causes particular outcomes to be biased depending on the input states.
Note that higher-order loss errors constitute an inherent violation of the Knill-Laflamme conditions, as expected since we are working with a continuous-variable code, but which can be suppressed with appropriate parameter choices.

The most obvious contribution to deformation is from the normalisation constant mis-match, as each index $S_{KJ}(n,m)$ is weighted by $(\mathcal{N}_J \underline{\mathcal{N}}_K)^{-1/2}$, which persists for all values of $\Gamma$.
However it is not the sole determiner of deformation, as, for instance, at $L=3$, $\alpha=2$, and $\Gamma=0$ we find
\begin{equation}
S(10,6)
=
\begin{pmatrix}
-\frac{0.36531}{\sqrt{\mathcal{N}_0 \underline{\mathcal{N}}_0}} & 0 \\
0 & -\frac{0.16605}{\sqrt{\mathcal{N}_1 \underline{\mathcal{N}}_1}}
\end{pmatrix},
\end{equation}
which shows deformation independent of the normalisation constants.
Regardless, we find that deformation overall is suppressed at sufficiently high amplitudes, where the contributions from weak Pauli measurement---and the sources themselves, such as normalisation mis-match---become small.
In this regime the transformations overall become approximately unitary.
We also propose in Sec.~\ref{sec:dedeformation} a probabilistic method for otherwise correcting deformation.

\subsection{Performance}
\label{sec:performance}

To quantify performance we use the channel fidelity, $F_C$ \cite{reimpell2005iterative}.
We iterate loss and correction $N$ times, wherein we apply the loss channel, $\mathcal{L}_{\gamma}$, with segment loss $\gamma = 1 - ( 1 - \Gamma )^{1/N}$, and correction, $\mathcal{C}$, to the first mode of an input entangled state
\begin{equation}
\rho_0 = \vert \tilde{\Phi}^+ \rangle \langle \tilde{\Phi}^+ \vert,
\quad
\vert \tilde{\Phi}^{+} \rangle
=
\frac{\vert \tilde{0} \tilde{0} \rangle + \vert \tilde{1} \tilde{1} \rangle}{\sqrt{2}},
\end{equation}
while applying the identity map, $\mathcal{I}$, to the other mode, iterating the channel as
\begin{equation}
\rho_{i+1}
=
\bigl(
\mathcal{C}
\otimes
\mathcal{I}
\bigr)
\bigl(
\bigl(
\mathcal{L}_{\gamma}
\otimes
\mathcal{I}
\bigr)
\bigl(
\rho_i
\bigr)
\bigr),
\label{eqn:channel_iteration}
\end{equation}
until $\rho_N$ is reached, then calculate the fidelity with the original state,
\begin{equation}
F_C
\equiv
\langle \tilde{\Phi}^+ \vert
\rho_N
\vert \tilde{\Phi}^+ \rangle.
\end{equation}
We take $\mathcal{C}$ to include both the teleportation protocol described earlier and the correction of logical $X$ and $Z$ errors indicated by the syndrome measurements, averaged over $(n,m)$ up to an appropriate cutoff.
In Appendix~\ref{app:in_out_map}, we derive an expression, Eq.~(\ref{eqn:rho_out}), to calculate the effects of the combined channel in Eq.~(\ref{eqn:channel_iteration}).

\begin{figure*}[t]
\includegraphics{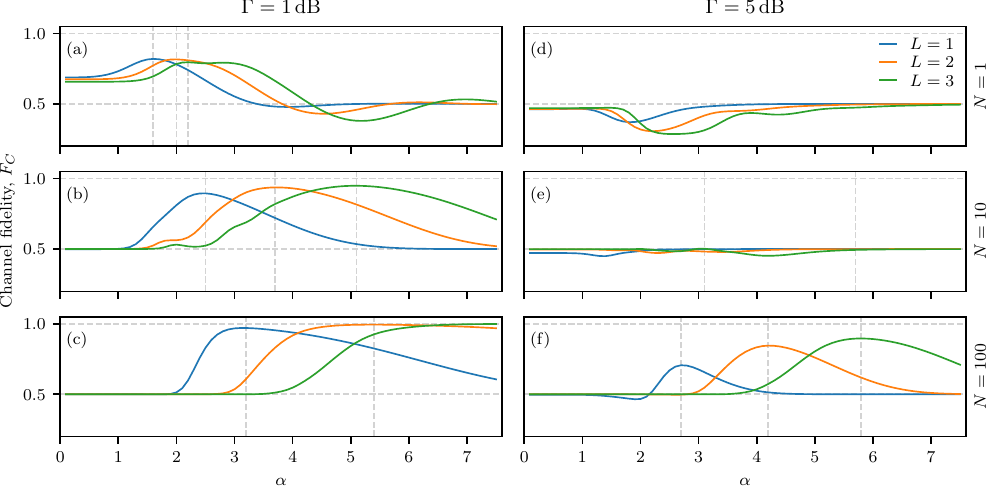}
\caption{\label{fig:channel_fidelity} The channel fidelity for different code orders at $1 \text{ dB}$ (left) and $5 \text{ dB}$ (right), shown for increasing numbers of channel iterations.
Peak fidelity is indicated with dashed vertical lines, except in cases where the optima are found numerically to be at the edge of the given span of $\alpha$.}
\end{figure*}

Fig.~\ref{fig:channel_fidelity} shows that lower code orders see an upward trend in performance earlier than higher orders, as we expect from the trends in normalisation constant mis-match.
However the lower orders do not generally reach the same performance levels if we continue higher in amplitude.
As seen in Ref.~\cite{hastrup2022all}, we observe that there is generally a maximum fidelity achieved at a given number of iterations, and that increasing the number of iterations also demands an increased amplitude to achieve the same fidelity.
We see generally that iteration number and state amplitude must both be increased to see performance improvements.
This is expected, since while smaller segments reduce the uncorrectable loss errors at each stage, each new teleportation risks deforming the state.

In Table~\ref{tab:f_target}, there is a superlinear decrease with respect to $L$ in the number of iterations required to achieve the target fidelities.
This comes at a cost in state amplitude, which need be higher with order, meaning that each iteration becomes more expensive, as not only do the photon-number measurements require higher resolutions, but the mean photon-number of the ancillae must also increase.
However, looking to $F_{\text{target}} = 0.999$, between $L=1$ and $L=3$ the required number of iterations is reduced by a factor of about $70.6$, while amplitude has only increased by about $1.9$, or about $3.6$ in terms of mean photon-number.

\begin{table}
\caption{\label{tab:f_target}
The minimum number of iterations, $N$, and associated amplitude, $\alpha$, required to achieve target fidelities, $F_{\text{target}}$, at $\Gamma = 1 \text{ dB}$ for different orders, $L$, of the cat code.
For a given number of iterations we scanned over $\alpha$ in increments of $\Delta \alpha = 0.1$ until the fidelity threshold was satisfied.
Note that the mean photon-number is approximately equal to the square of the amplitude in this regime.}
\begin{ruledtabular}
\begin{tabular}{ccccccc}
&\multicolumn{2}{c}{$L=1$}&\multicolumn{2}{c}{$L=2$}&\multicolumn{2}{c}{$L=3$}\\
$F_{\text{target}}$ & $N$ & $\alpha$ & $N$ & $\alpha$ & $N$ & $\alpha$ \\
\hline
$ 0.95$& $   44$ & $ 2.9$ & $ 14$ & $ 3.8$ & $ 11$ & $ 5.0$ \\
$ 0.99$& $  500$ & $ 3.6$ & $ 68$ & $ 5.1$ & $ 38$ & $ 6.4$ \\
$0.999$& $10022$ & $ 4.3$ & $423$ & $ 6.4$ & $142$ & $ 8.2$
\end{tabular}
\end{ruledtabular}
\end{table}

\section{Probabilistic deformation correction} \label{sec:dedeformation}

The deformations described in Sec.~\ref{sec:syndromes} are detectable, but, as they are non-unitary, cannot be corrected deterministically.
Here we will show that the same teleportation scheme as for loss-correction can be used to probabilistically correct deformation, provided one can finely tune the ancilla.
Because deformation is normally only suppressed by working in a higher-amplitude regime---exacerbating loss errors---correcting deformation explicitly allows one to achieve optimal performance at lower amplitudes, both reducing the resource cost during transmission and improving fidelity.

In Table~\ref{tab:deformation_correction} we see the differences in fidelity supposing one is able to perfectly implement the inverse of all accumulated deformations during transmission.
We see an improvement over the optimal performance achieved exclusively via Pauli-correction, with optima appearing at lower amplitudes.
However, the incidence of transmission failure increases, particularly for the lower-order codes, as these failures are generally associated with lower photon-number measurements.

The falling off in performance at excessively low amplitudes is likely due to extreme deformations exacerbating loss errors, as even if those contributions are made smaller by the reduced amplitude, they will be multiplied in correcting the deformation.
Similarly, since deformation technically does not impact $\vert \tilde{0} \rangle$ and $\vert \tilde{1} \rangle$, but higher orders of loss can contribute to these states during transmission, in inverting the deformation these loss-error contributions can be exacerbated, worsening fidelity.

\begin{table}[h]
\caption{\label{tab:deformation_correction}
Performance statistics observed over $1000$ simulations of the channel with $N=10$ iterations and loss $\Gamma=1\text{ dB}$, averaged over inputs of the six Pauli eigenstates, with amplitudes at fractions of an optimal value, $\alpha_{\text{opt}}$, derived from Fig.~\ref{fig:MEP_compare}.
$\Delta F$ is the percent difference in average fidelity between deformation-corrected and Pauli-corrected output states, $\Delta F = F_{\text{DC}} - F_{\text{PC}}$.
$\Delta F_{\text{opt}}$ is the percent difference between the average deformation-corrected output state and the average Pauli-corrected state at optimal amplitude, $\Delta F_{\text{opt}} = F_{\text{DC}} - F_{\text{opt}}$.
Only successful transmissions are accounted for in the average fidelities.
The percentage of failures witnessed throughout each sample is listed.}
\begin{ruledtabular}
\begin{tabular}{cccccccccc}
& \multicolumn{3}{c}{$L=1$} & \multicolumn{3}{c}{$L=2$} & \multicolumn{3}{c}{$L=3$}\\
& $\Delta F$ & $\Delta F_{\text{opt}}$\footnotemark[1] & $\%_{\text{fail}}$ & $\Delta F$ & $\Delta F_{\text{opt}}$\footnotemark[1] & $\%_{\text{fail}}$ & $\Delta F$ & $\Delta F_{\text{opt}}$\footnotemark[1] & $\%_{\text{fail}}$ \\
\hline
$\alpha_{\text{opt}}$\footnote{$\alpha_{\text{opt}} = \{2.5,3.7,5.1\}$, $F_{\text{opt}}=\{93.0,96.5,96.5\}$ for $L=\{1,2,3\}$} & $ 0.9$ & $+0.9$ & $ 0.1$ & $ 0.8$ & $+0.8$ & $ 1.8$ & $ 1.1$ & $+1.1$ & $ 0.0$ \\
$         \times 0.9$ & $ 3.0$ & $+1.9$ & $ 0.2$ & $ 2.5$ & $+1.8$ & $ 3.2$ & $ 2.4$ & $+1.5$ & $ 0.0$ \\
$         \times 0.8$ & $ 6.6$ & $+1.3$ & $ 2.2$ & $ 6.4$ & $+2.0$ & $ 8.9$ & $ 6.4$ & $+2.1$ & $ 0.0$ \\
$         \times 0.7$ & $10.3$ & $+0.7$ & $12.4$ & $11.2$ & $-0.6$ & $15.3$ & $12.6$ & $+2.2$ & $ 0.0$ \\
$         \times 0.6$ & $12.8$ & $+1.0$ & $47.2$ & $13.9$ & $-8.7$ & $17.7$ & $18.0$ & $-0.7$ & $ 0.0$ \\
$         \times 0.5$ & $13.0$ & $-5.1$ & $79.7$ & $16.3$ & $-6.5$ & $34.3$ & $12.3$ & $-15.5$ & $ 0.4$ \\
\end{tabular}
\end{ruledtabular}
\end{table}

\subsection{Ancilla-biasing and code-switching}
\label{sec:ancilla_tricks}

To implement the inverse deformation optically, one can bias the ancilla in the teleporter, where with a bias $x$ introduced as
\begin{equation}
\vert \tilde{\Phi}^{(x)} \rangle
=
\frac{\vert \underline{\tilde{0}} \tilde{0} \rangle + x \vert \underline{\tilde{1}} \tilde{1} \rangle}{\sqrt{1 + \vert x \vert^2}},
\label{eqn:ancilla_bias}
\end{equation}
the syndromes are transformed as follows:
\begin{equation}
S^{(x)} (n,m) \propto
\begin{pmatrix}
1 & 1 \\
x & x
\end{pmatrix}
S(n,m),
\label{eqn:transform_bias}
\end{equation}
as $x$ becomes associated with the index $K=1$ in Eq.~(\ref{eqn:syndrome}).
With knowledge of the deformations corresponding to each $S(n,m)$, this transformation can therefore produce a set containing an arbitrary deformation of one's choice.

To assess the actual probability of selecting a desired transformation, one must consider the total probability of each outcome, $S(n,m)$, of which many are---up to a constant---degenerate.
For $L=1$ there are considerably fewer distinct deformations, meaning that if we need a particular transformation then it is statistically favourable to work at this code order.
This may seem prohibitive, considering the performance gains of working at higher orders during transmission, but it is in fact possible to switch into a different basis supposing we can prepare the appropriate ancilla.
The basis one teleports into is theoretically arbitrary, as one could prepare an ancilla
\begin{equation}
\vert \tilde{\Phi}^{+}_{(L,L')} \rangle
=
\frac{\vert \underline{\tilde{0}}_L \tilde{0}_{L'} \rangle + \vert \underline{\tilde{1}}_L \tilde{1}_{L'} \rangle}{\sqrt{2}},
\end{equation}
which in teleportation would take the input state from code order $L$ to $L'$, although this initial code-switching teleportation would have the set of syndromes $S$ corresponding to $L$.
One could similarly use this method to teleport into a cat basis of different amplitude, or into the basis of another code entirely.

In total, to implement our deformation-correction scheme, we propose two additional teleportations at the terminus of transmission: one to change into an $L=1$ cat code of sufficient amplitude to be in the unitary regime; a second with an ancilla of bias $x$ whose most probable outcome has a deformation proportional to the inverse of the post-transmission deformation.
The post-transmission deformation is calculated from the matrix product of the syndromes associated with each photon-number measurement pair made during transmission.
We will assume one has access to $X$ and $Z$ gates, although only $Z$ is accessible through Gaussian operations.

Using what we have described, in Table~\ref{tab:bias_probs} we see the success probability for selecting our desired transformation at different levels of bias.
For no bias, $x=1$, the probability is near $100\%$ to select for a deformation-less output on account of the transformations in this regime being close to unitary.
The sudden change to an average of $50\%$ success probability is on account of Eq.~(\ref{eqn:transform_bias}), which produces diagonal transformations with deformation ratios
\begin{equation}
d_{\text{diag}}
\propto
\frac{1}{x}
\label{eqn:bias_diag}
\end{equation}
while off-diagonal transformations have deformation ratios
\begin{equation}
d_{\text{flip}}
\propto
x
\label{eqn:bias_flip}
\end{equation}
meaning that with bias our formerly unitary bit-flips now encode the inverse deformation compared to the other transformations.
In essence, in the unitary regime, if our bias produces diagonal transformations proportional to a matrix $D$, then our total set of transformations will be approximately
\begin{equation}
\Bigl\{
D,
X \bigl( D^{-1} \bigr)
\Bigr\},
\end{equation}
up to some additional $Z$ which can be corrected deterministically via rotation.

In Table~\ref{tab:bias_probs}, because the effect that we are taking advantage of is weak Pauli $Z$ measurement, only the eigenstates $\vert \tilde{0} \rangle$ and $\vert \tilde{1} \rangle$ have their probabilities affected by bias, where we see that the probability of success is dramatically suppressed for $\vert \tilde{0} \rangle$.
This effect on probability would occur for any input state where the probabilistic balance of the codewords is not equal.
As the deformation this would correct is related to the amount of information revealed about our initial state, and this technique uses the same mechanism to effectively produce a result contradictory to the information previously revealed, it is thus naturally unfavourable.
Further, given the effect outlined via Eqs.~(\ref{eqn:bias_diag}) and (\ref{eqn:bias_flip}), where off-diagonal transformations encode the inverse deformation relative to diagonal transformations, each attempt to correct deformation this way risks worsening the deformation by a factor of two, which will also further suppress $p_{\tilde{0}}$ or $p_{\tilde{1}}$.

\begin{table}[t!]
\caption{\label{tab:bias_probs}
Probability of an outcome corresponding to the desired deformation up to second-decimal precision after a bias, $x$, in Eq.~(\ref{eqn:ancilla_bias}).
The probabilities are averaged across the Pauli eigenstates to produce $p_{\text{avg}}$, while $p_{\tilde{0}}$ and $p_{\tilde{1}}$ are for $\vert \tilde{0} \rangle$ and $\vert \tilde{1} \rangle$ individually.
$X$ and $Z$ errors have been ignored in selecting desired deformations.
The maximum bias is chosen based on the average deformation ratio corresponding to $L=3$ and $\alpha = 0.7 \times \alpha_{\text{opt}}$ in Table.~\ref{tab:deformation_correction}, where $d \approx 4.76$.
These calculations were performed for $L=1$, $\alpha=4$, and $\Gamma = 0$.}
\begin{ruledtabular}
\begin{tabular}{cccc}
bias ($x$) & $p_{\text{avg}}$ & $p_{\tilde{0}}$ & $p_{\tilde{1}}$ \\
\hline
$  1$ & $99.9$ & $99.9$ & $99.9$ \\
$1/2$ & $ 0.5$ & $20.0$ & $80.0$ \\
$1/3$ & $ 0.5$ & $10.0$ & $90.0$ \\
$1/4$ & $ 0.5$ & $ 5.9$ & $94.1$ \\
$1/5$ & $ 0.5$ & $ 3.8$ & $96.2$ \\
\end{tabular}
\end{ruledtabular}
\end{table}

\section{Conclusion}
\label{sec:conclusion}

We have generalised the protocol proposed in Ref.~\cite{hastrup2022all} to higher orders of the cat code, and found that these higher orders can be used to significantly reduce the number of iterations required to achieve target performance.
The performance improvements, however, come at a cost in state amplitude, which makes the protocol's photon-number measurements more demanding, and would also make ancilla generation more expensive.
The mean photon number has been used as a proxy for state cost previously \cite{albert2018performance,grimsmo2020quantum}, and the high-distance Fock structure of higher-order cat codes may also impact the rate and fidelity with which they can be generated.

Within our choice of basis, we have also identified codeword normalisation mis-match as an inherent source of error in the teleportation scheme, which is complementary to the codeword overlap seen in the basis used in the original proposal.
This normalisation mis-match contributes to weak Pauli $Z$ measurements in the protocol, but we have found that the effect is minimised at sufficient amplitudes, and moreover we are able to identify the deformation via our syndrome measurements.
Further, our proposal for a probabilistic scheme to correct deformation shows that even if impractical to implement, deformation is technically reversible.
It also highlights two interesting properties of the teleporter:
that it can produce a set of two arbitrary but related diagonal and off-diagonal transformations, and that it can be used to change the basis of our state.
Both of these techniques would require very robust ancilla generation.

Ultimately, we find that if one can realise the measurements and state generation required to teleport higher-order cat states, then transmission will require far fewer correction stages.

The results of this paper can be replicated using the Python project \texttt{CatTelecorrection} \cite{Boon_CatTelecorrection_2026}, which is our numerical implementation of the work presented here.

\begin{acknowledgments}
A.J.B. and N.Q. acknowledge the support from the Ministère de l'Économie et de l'Innovation du Québec, and the Natural Sciences and Engineering Research Council of Canada.
Our research was enabled in part by support provided by the Institut Transdisciplinaire d'Information Quantique (INTRIQ), a strategic cluster funded by the Fonds de recherche du Québec.
We also thank Javier Martínez-Cifuentes for his insights when writing this paper, and Jacob Hastrup for valuable discussions as well as for sharing his numerical results from his $L=1$ cat telecorrection protocol.
\end{acknowledgments}

\appendix

\begin{widetext}
\section{Choice of basis} \label{app:bases}

The basis of cat codewords we chose to use in this work differs from the choice originally proposed in Ref.~\cite{hastrup2022all}.
We will distinguish the two bases by the effect of the rotation
\begin{equation}
\hat{R}_L
=
\exp
\left(
\frac{
i \pi 
\hat{a}^{\dagger} \hat{a}}{
L+1}
\right),
\label{eqn:rotation}
\end{equation}
where in our choice of basis the effect of $\hat{R}_L$ is a logical $Z$, and for the basis in Ref.~\cite{hastrup2022all} it is a logical $X$.
On account of this, we will refer to the bases as $R_Z$ cats,
\begin{equation}
\vert \tilde{0} \rangle_Z
=
\frac{1}{\sqrt{\mathcal{N}_0}}
\sum^{2 L + 1}_{m=0}
\vert e^{m \pi i / (L + 1)} \alpha \rangle,
\quad
\vert \tilde{1} \rangle_Z
=
\frac{1}{\sqrt{\mathcal{N}_1}}
\sum^{2 L + 1}_{m=0}
(-1)^m
\vert e^{m \pi i / (L + 1)} \alpha \rangle,
\end{equation}
and $R_X$ cats,
\begin{equation}
\vert \tilde{0} \rangle_X
=
\frac{1}{\sqrt{\mathcal{N}_X}}
\sum^{L}_{m=0}
\vert e^{2 m \pi i / (L + 1)} \alpha \rangle,
\quad
\vert \tilde{1} \rangle_X
=
\frac{1}{\sqrt{\mathcal{N}_X}}
\sum^{L}_{m=0}
\vert e^{(2 m + 1) \pi i / (L + 1)} \alpha \rangle,
\label{eqn:Rx_cats}
\end{equation}
respectively.
The Pauli $X$ eigenstates in the $R_X$ basis are equivalent to Pauli $Z$ eigenstates in $R_Z$,
\begin{equation}
\vert \tilde{\pm} \rangle_X
=
\vert \tilde{0} / \tilde{1} \rangle_Z,
\label{eqn:EigEquiv}
\end{equation}
but there are key differences:
in $R_Z$ the codewords are perfectly orthogonal, $\langle \tilde{0} \vert \tilde{1} \rangle_Z = 0$, but have different normalisation constants, $\mathcal{N}_0 \neq \mathcal{N}_1$;
in $R_X$ the codewords are not perfectly orthogonal, $\langle \tilde{0} \vert \tilde{1} \rangle_X \neq 0$, but have a common normalisation constant, $\mathcal{N}_X$.

These differences have been commented on in Ref.~\cite{bergmann2016quantum}, where they are noted to lead to practical differences in performance against the loss channel, as normalisation mis-match leads to deformation in correcting the code, while non-orthogonality means an inherent level of indistinguishability between the codewords---both distinct violations of the Knill-Laflamme conditions.
However, both sources of error are noted to diminish as $\alpha \rightarrow \infty$, and we show here how the two are intrinsically linked, and show for $L=1$ that their impact on performance is relegated to the sub-optimal regime.

To begin with, an $R_X$ qubit requires additional normalisation to account for overlap \cite{bergmann2016quantum},
\begin{equation}
\vert \tilde{\Psi} \rangle_X
=
\frac{ \mu \vert \tilde{0} \rangle_X + \nu \vert \tilde{1} \rangle_X }{\sqrt{ 1 + 2 {\text{Re}}{( \mu \nu^* )} \langle \tilde{0} \vert \tilde{1} \rangle_X }},
\end{equation}
where we take $\alpha \in \mathbb{R}$ and so $\langle \tilde{0} \vert \tilde{1} \rangle_X \in \mathbb{R}$.
With this, we can define the properly normalised $R_X$ Pauli $X$ eigenstates,
\begin{equation}
\vert \tilde{\pm} \rangle_X
=
\frac{ \vert \tilde{0} \rangle_X \pm \vert \tilde{1} \rangle_X }{\sqrt{ 1 \pm \langle \tilde{0} \vert \tilde{1} \rangle_X }},
\label{eqn:Rx_eig_plus}
\end{equation}
and expand the numerator as follows:
\begin{align}
\vert \tilde{0} \rangle_X \pm \vert \tilde{1} \rangle_X
&=
\sum^{L}_{m=0}
\frac{
\vert e^{2 m \pi i / (L + 1)} \alpha \rangle
\pm
\vert e^{(2 m + 1) \pi i / (L + 1)} \alpha \rangle}{\sqrt{\mathcal{N}_X}}
\\
&=
\sum^{2 L + 1}_{m=0}
\frac{( \pm 1 )^m
\vert e^{ m \pi i / (L + 1)} \alpha \rangle}{\sqrt{\mathcal{N}_X}}
.
\label{eqn:OrderEquiv}
\end{align}
Now, using this and Eq.~(\ref{eqn:Rx_eig_plus}) on the LHS of Eq.~(\ref{eqn:EigEquiv}), we find
\begin{align}
\sum^{2 L + 1}_{m=0}
\frac{
( \pm 1 )^m
\vert e^{ m \pi i / (L + 1) } \alpha \rangle}{\sqrt{ \mathcal{N}_X \bigl( 1 \pm \langle \tilde{0} \vert \tilde{1} \rangle_X \bigr) }}
&=
\sum^{2 L + 1}_{m=0}
\frac{( \pm 1 )^m
\vert e^{ m \pi i / (L + 1)} \alpha \rangle}{\sqrt{\mathcal{N}_{0/1}}}
\\
\mathcal{N}_X
\bigl( 1 \pm \langle \tilde{0} \vert \tilde{1} \rangle_X \bigr)
&=
\mathcal{N}_{0/1},
\end{align}
showing that the source of intrinsic error in one basis is directly related to that of the other, and with which we write
\begin{equation}
\frac{\mathcal{N}_0}{\mathcal{N}_1}
=
\frac{ 1 + \langle \tilde{0} \vert \tilde{1} \rangle_X }{ 1 - \langle \tilde{0} \vert \tilde{1} \rangle_X }.
\end{equation}
Thus, we see that as codeword overlap in the $R_X$ basis grows small, $\langle \tilde{0} \vert \tilde{1} \rangle_X \rightarrow 0$, the ratio between $R_Z$ normalisation constants approaches unity, $\mathcal{N}_0 / \mathcal{N}_1 \rightarrow 1$.
However, as the form of these errors is distinct, we must consider how they impact performance.

\begin{figure*}[t]
\includegraphics{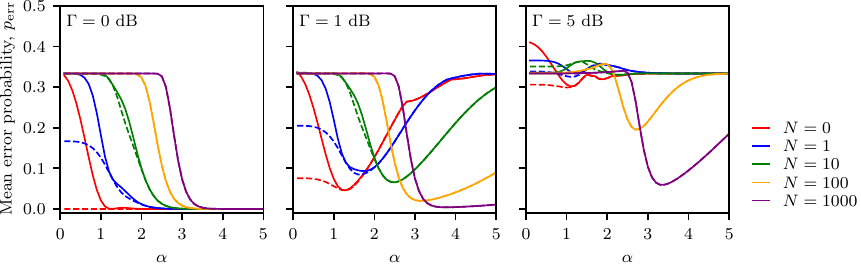}
\caption{\label{fig:MEP_compare} A comparison of mean error probabilities between the $R_Z$ (dashed; our work) and $R_X$ (solid; Ref.~\cite{hastrup2022all}), calculated after $N$ applications of loss and correction then a single additional uncorrected application of loss, performed at order $L=1$.
For $N=0$ no correction is applied.}
\end{figure*}

Numerically, we will make a direct comparison to the results from Ref.~\cite{hastrup2022all}, mimicking the original proposal's choice of the mean error probability as a metric, its use of an additional segment of loss, and its choice of code order $L=1$.
The mean error probability is calculated by averaging over the trace distance of each pair of Pauli eigenstates---$\{\vert \tilde{0} \rangle, \vert \tilde{1} \rangle \}$,
$\{\vert \tilde{+} \rangle, \vert \tilde{-} \rangle \}$, and
$\{\vert \tilde{+i} \rangle, \vert \tilde{-i} \rangle \}$---after they have been passed through the channel, iterating as $\rho_{i+1} = \mathcal{C}(\mathcal{L}_{\gamma}(\rho_i))$ with $\rho_0$ an eigenstate, where for a given pair of Pauli eigenstates, \textit{e.g.}, $\rho^{X+}_0 = \vert \tilde{+} \rangle \langle \tilde{+} \vert$ and $\rho^{X-}_0 = \vert \tilde{-} \rangle \langle \tilde{-} \vert$, we have
\begin{equation}
p^X_{\text{err}}
=
\frac{1}{2}
-
\frac{1}{4}
\bigl| \bigl| 
\mathcal{L}_{\gamma}(\rho^{X+}_N)
-
\mathcal{L}_{\gamma}(\rho^{X-}_N)
\bigr| \bigr|_1,
\end{equation}
with $\gamma = 1 - (1 - \Gamma)^{1/(N+1)}$, and the average is given by
\begin{equation}
p_{\text{err}}
\equiv
\frac{p^X_{\text{err}} + p^Y_{\text{err}} + p^Z_{\text{err}}}{3}.
\end{equation}
In Fig.~\ref{fig:MEP_compare} we compare performance in $R_Z$ and $R_X$ this way, and see that there are only significant differences in the sub-break-even regime for $L=1$.
As $\alpha$ increases, sources of error in both $R_Z$ and $R_X$ are minimised, and the bases become approximately identical in performance.

\section{Generalised expressions for cat telecorrection} \label{app:teleportation}

\subsection{Mapping of input logical states to output logical states}
\label{app:in_out_map}

We will begin with the effect of the loss channel on a coherent state.
Using the Kraus operators of the loss channel given in Eq.~(\ref{eqn:loss_kraus}), for a coherent state of amplitude $\alpha$ and phase $\phi$, we find
\begin{align}
\hat{B}^{(l)}_{\Gamma}
\vert e^{i\phi}\alpha \rangle
&=
\sqrt{\frac{\Gamma}{1-\Gamma}}^{\,l}
\hat{a}^l \, \sqrt{1 - \Gamma}^{\, \hat{n}} \vert e^{i\phi}\alpha \rangle
\\
&=
\sqrt{\frac{\Gamma}{1-\Gamma}}^{\,l}
\hat{a}^l \, e^{-\Gamma \alpha^2/2} \vert e^{i\phi} \sqrt{1 - \Gamma} \alpha \rangle
\\
&=
\sqrt{\frac{\Gamma}{1-\Gamma}}^{\,l}
e^{-\Gamma \alpha^2/2} \sqrt{1 - \Gamma}^{\,l} e^{il\phi} \alpha^l \vert e^{i\phi} \sqrt{1 - \Gamma} \alpha \rangle
\\
&=
e^{-\Gamma \alpha^2/2}
\bigl(
\sqrt{\Gamma} e^{i\phi} \alpha
\bigr)^l
\vert e^{i\phi} \sqrt{1 - \Gamma} \alpha \rangle.
\label{eqn:coherent_loss}
\end{align}
Going forward we will denote the coherent states composing a particular order of cat code as follows,
\begin{equation}
\underline{\alpha}_j \equiv e^{i j \pi / (L+1)} \sqrt{1 - \Gamma} \alpha,
\end{equation}
where an underbar has been used to indicate amplitude damping, $\underline{\alpha} \equiv \sqrt{1-\Gamma} \alpha$, and will also be used to indicate logical codewords, $\vert \underline{\tilde{J}} \rangle$, and normalisation constants, $\underline{\mathcal{N}}_J$, belonging to states prepared at a damped amplitude.

For a state in the cat basis described by Eqs.~(\ref{eqn:codewords_coherent0}) and (\ref{eqn:codewords_coherent1}), we can write
\begin{equation}
\rho_{\text{in}}
=
\sum^{\{0,1\}}_{J,J'}
q_{JJ'}
\vert \tilde{J} \rangle
\langle \tilde{J}' \vert
\\
=
\sum^{\{0,1\}}_{J,J'}
\sum^{2 L + 1}_{j,j'=0}
q_{JJ'}
\frac{(-1)^{jJ + j'J'}}{\sqrt{\mathcal{N}_J \mathcal{N}_{J'}}}
\vert \alpha_j \rangle
\langle \alpha_{j'} \vert,
\label{eqn:cat}
\end{equation}
which will also describe any state immediately following teleportation, as here teleportation restores the state to the original basis.
Now, for our cat state of Eq.~(\ref{eqn:cat}), the effects of loss are as follows:
\begin{align}
\mathcal{L}_{\Gamma}(\rho_{\text{in}})
&=
\sum^{\infty}_{l=0}
\hat{B}^{(l)}_{\Gamma}
\rho_{\text{in}}
\left[\hat{B}^{(l)}_{\Gamma}\right]^{\dagger}
\\
&=
\sum^{\infty}_{l=0}
\sum^{\{0,1\}}_{J,J'}
q_{J J'}
\hat{B}^{(l)}_{\Gamma}
\vert \tilde{J} \rangle
\langle \tilde{J'} \vert
\left[\hat{B}^{(l)}_{\Gamma}\right]^{\dagger}
\\
&=
\sum^{\infty}_{l=0}
\sum^{\{0,1\}}_{J,J'}
\sum^{2L+1}_{j,j'=0}
q_{J J'}
\frac{(-1)^{jJ+j'J'}}{\sqrt{\mathcal{N}_J \mathcal{N}_{J'}}}
\hat{B}^{(l)}_{\Gamma}
\vert \alpha_j \rangle
\langle \alpha_{j'} \vert
\left[\hat{B}^{(l)}_{\Gamma}\right]^{\dagger}
\\
&=
\sum^{\infty}_{l=0}
\sum^{\{0,1\}}_{J,J'}
\sum^{2L+1}_{j,j'=0}
q_{J J'}
\frac{(-1)^{jJ+j'J'}}{\sqrt{\mathcal{N}_J \mathcal{N}_{J'}}}
\frac{\left[ \Gamma e^{i(j-j')\pi/(L+1)} \alpha^2 \right]^l}{l!} 
e^{-\Gamma \alpha^2}
\vert \underline{\alpha}_j \rangle
\langle \underline{\alpha}_{j'} \vert
\label{eqn:lossy_cat_discrete}
\\
&=
\sum^{\{0,1\}}_{J,J'}
\sum^{2L+1}_{j,j'=0}
q_{J J'}
\frac{(-1)^{jJ+j'J'}}{\sqrt{\mathcal{N}_J \mathcal{N}_{J'}}}
{\exp}{\left[ \Gamma \alpha^2 e^{i (j-j') \pi /(L+1)} \right]}
e^{-\Gamma \alpha^2}
\vert \underline{\alpha}_j \rangle
\langle \underline{\alpha}_{j'} \vert,
\label{eqn:lossy_cat_full}
\end{align}
where from (\ref{eqn:lossy_cat_discrete}) to (\ref{eqn:lossy_cat_full}) the sum over each order of loss has been re-written as
\begin{equation}
\sum^{\infty}_{l=0}
\frac{\left[ \Gamma e^{i(j-j')\pi/(L+1)} \alpha^2 \right]^l}{l!}
=
{\exp}{\left[ \Gamma \alpha^2 e^{i (j-j') \pi /(L+1)} \right]}.
\end{equation}

The lossy cat state is then mixed via a ${50}{:}{50}$ beamsplitter with the first mode of an encoded $\Phi^+$ ancilla given by
\begin{equation}
\rho_{\Phi}
=
\frac{1}{2}
\sum^{\{0,1\}}_{K,K'}
\vert \underline{\tilde{K}} \tilde{K} \rangle
\langle \underline{\tilde{K'}} \tilde{K'} \vert,
\label{eqn:ancilla}
\end{equation}
where the ancilla is composed of logical cat states, and has its first mode prepared at a reduced amplitude to match that of the lossy input state.
The action of a real symmetric ${50}{:}{50}$ beamsplitter on two modes containing coherent states is given by
\begin{equation}
\hat{U}
\vert
\alpha, \, \beta
\rangle
=
\biggl|  
\frac{\alpha+\beta}{\sqrt{2}},
\,
\frac{\beta-\alpha}{\sqrt{2}}
\biggr>.
\end{equation}
Using this we find
\begin{align}
\rho_M
&=
\hat{U}_{1,2}
\left[
\mathcal{L}(\rho_{\text{in}})
\otimes
\rho_{\Phi}
\right]
\hat{U}^{\dagger}_{1,2}
\\
&=
\sum^{\{0,1\}}_{\substack{J,J', \\ K,K'}}
\sum^{2L+1}_{j,j'=0}
q_{JJ'}
\frac{
(-1)^{jJ + j'J'}}{
\sqrt{\mathcal{N}_J \mathcal{N}_{J'}}}
{\exp}{\left[ \Gamma \alpha^2 e^{i (j-j') \pi /(L+1)} \right]}
e^{-\Gamma \alpha^2}
\hat{U}_{1,2}
\vert \underline{\alpha}_j \, \underline{\tilde{K}} \, \tilde{K} \rangle
\langle \underline{\alpha}_{j'} \, \underline{\tilde{K}'} \tilde{K'} \vert
\hat{U}^{\dagger}_{1,2}
\\
&=
\sum^{\{0,1\}}_{\substack{J,J', \\ K,K'}}
\sum^{2L+1}_{\substack{j,j', \\ k,k' =0}}
q_{JJ'}
\frac{
(-1)^{jJ + j'J' + kK + k'K'}}{
\sqrt{\mathcal{N}_J \underline{\mathcal{N}}_K \mathcal{N}_{J'} \underline{\mathcal{N}}_{K'}}}
{\exp}{\left[ \Gamma \alpha^2 e^{i (j-j') \pi /(L+1)} \right]}
e^{-\Gamma \alpha^2}
\hat{U}_{1,2}
\vert \underline{\alpha}_j \, \underline{\alpha}_k \, \tilde{K} \rangle
\langle \underline{\alpha}_{j'} \, \underline{\alpha}_{k'} \tilde{K'} \vert
\hat{U}^{\dagger}_{1,2}
\\
&=
\sum^{\{0,1\}}_{\substack{J,J', \\ K,K'}}
\sum^{2L+1}_{\substack{j,j', \\ k,k' =0}}
q_{JJ'}
\frac{
(-1)^{jJ + j'J' + kK + k'K'}}{
\sqrt{\mathcal{N}_J \underline{\mathcal{N}}_K \mathcal{N}_{J'} \underline{\mathcal{N}}_{K'}}}
{\exp}{\left[ \Gamma \alpha^2 e^{i (j-j') \pi /(L+1)} \right]}
e^{-\Gamma \alpha^2}
\vert \underline{\gamma}_{(j,k)} \, \underline{\beta}_{(k,j)} \, \tilde{K} \rangle
\langle \underline{\gamma}_{(j',k')} \, \underline{\beta}_{(k',j')} \tilde{K'} \vert,
\end{align}
where we have used
\begin{align}
\underline{\gamma}_{(j,k)}
&\equiv
\left[
e^{i j \pi /(L+1)}
+
e^{i k \pi /(L+1)}
\right]
\sqrt{1-\Gamma} \alpha / \sqrt{2},
\label{eqn:gamma}
\\
\underline{\beta}_{(k,j)}
&\equiv
\left[
e^{i k \pi /(L+1)}
-
e^{i j \pi /(L+1)}
\right]
\sqrt{1-\Gamma} \alpha / \sqrt{2}.
\label{eqn:beta}
\end{align}
to simplify the mixed output of the beamsplitter; one can see here the importance of matching the amplitude of the ancilla's first mode to that of the lossy transmitted state, as otherwise the pairs of exponentials in either of Eqs.~(\ref{eqn:gamma}) or (\ref{eqn:beta}) would lack a common factor.

Next, two photon-number measurements are made on the first two modes: that of the input state and the reduced-amplitude mode of the ancilla.
For coherent states a photon-number measurement is as follows for $\beta \in \mathbb{C}$:
\begin{equation}
\langle
n
\vert
\beta
\rangle
=
e^{-\vert \beta \vert^2/2}
\frac{\beta^n}{\sqrt{n!}}.
\label{eqn:count}
\end{equation}
Here, for brevity, we show the effect of the measurements only on the left bra-ket pair:
\begin{align*}
\langle n \, m \,
\vert \underline{\gamma}_{(j,k)} \, \underline{\beta}_{(k,j)} \rangle
&=
\frac{1}{\sqrt{n!m!}}
\,
{\exp}{\left[ - \frac{1}{2} \left( \vert \underline{\gamma}_{(j,k)} \vert^2 + \vert \underline{\beta}_{(k,j)} \vert^2 \right) \right]}
\left[\underline{\gamma}_{(j,k)}\right]^n
\left[\underline{\beta}_{(k,j)}\right]^m,
\end{align*}
where we find that the exponential is a constant,
\begin{align}
\vert \gamma_{(j,k)} \vert^2 + \vert \beta_{(k,j)} \vert^2
&=
\frac{\alpha^2}{2}
\left(
\bigl|
e^{i j \pi /(L+1)}
+
e^{i k \pi /(L+1)}
\bigr|^2
+
\bigl|
e^{i k \pi /(L+1)}
-
e^{i j \pi /(L+1)}
\bigr|^2
\right)
\notag
\\
&=
\frac{\alpha^2}{2}
\Bigl[
\left(
e^{i j \pi /(L+1)}
+
e^{i k \pi /(L+1)}
\right)
\left(
e^{-i j \pi /(L+1)}
+
e^{-i k \pi /(L+1)}
\right)
\Bigr.
\notag
\\
&\hphantom{=}+
\Bigl.
\left(
e^{i k \pi /(L+1)}
-
e^{i j \pi /(L+1)}
\right)
\left(
e^{-i k \pi /(L+1)}
-
e^{-i j \pi /(L+1)}
\right)
\Bigr]
\notag
\\
&=
\frac{\alpha^2}{2}
\Bigl[
4e^0
+
e^{i(j-k)\pi/(L+1)}
-
e^{i(j-k)\pi/(L+1)}
+
e^{i(k-j)\pi/(L+1)}
-
e^{i(k-j)\pi/(L+1)}
\Bigr]
\notag
\\
&=
2\alpha^2.
\end{align}
Thus we have
\begin{equation}
\langle n \, m \,
\vert \underline{\gamma}_{(j,k)} \, \underline{\beta}_{(k,j)} \rangle
=
\frac{e^{- (1-\Gamma) \alpha^2}}{\sqrt{n!m!}}
\left[
\sqrt{\frac{1-\Gamma}{2}} \alpha
\right]^{n+m}
\Bigl[ e^{i j \pi /(L+1)} + e^{i k \pi /(L+1)} \Bigr]^n
\Bigl[ e^{i k \pi /(L+1)} - e^{i j \pi /(L+1)} \Bigr]^m.
\end{equation}
Using this, the expression for the final state corresponding to a syndrome measurement $(n,m)$ is
\begin{align}
\rho_{\text{out}}(n,m)
&=
\bigl( \langle n \, m \vert \otimes \mathcal{I} \bigr)
\rho_M
\bigl( \vert n \, m \rangle \otimes \mathcal{I} \bigr)
\\
&=
\sum^{\{0,1\}}_{\substack{J,J', \\ K,K'}}
\sum^{2 L + 1}_{\substack{j,j', \\ k,k' =0}}
q_{JJ'}
\frac{
(-1)^{jJ + j'J' + kK + k'K'}}{
\sqrt{\mathcal{N}_J \underline{\mathcal{N}}_K \mathcal{N}_{J'} \underline{\mathcal{N}}_{K'}}}
\frac{e^{ (\Gamma - 2) \alpha^2}}{n!m!}
\left[
\frac{(1-\Gamma)}{2} \alpha^2
\right]^{n+m}
\notag
\\
&\hspace{2.2cm} \cdot
\Bigl[ e^{i j \pi / (L + 1)} + e^{i k \pi / (L + 1)} \Bigr]^n
\Bigl[ e^{i k \pi / (L + 1)} - e^{i j \pi / (L + 1)} \Bigr]^m
\notag
\\
&\hspace{2.2cm} \cdot
\Bigl[ e^{-i j' \pi / (L + 1)} + e^{-i k' \pi / (L + 1)} \Bigr]^n
\Bigl[ e^{-i k' \pi / (L + 1)} - e^{-i j' \pi / (L + 1)} \Bigr]^m
\notag
\\
&\hspace{2.2cm} \cdot
{\exp}{\left[ \Gamma \alpha^2 e^{i (j-j') \pi / (L + 1)} \right]}
\notag
\\
&\hspace{2.2cm} \cdot
\vert \tilde{K} \rangle
\langle \tilde{K'} \vert.
\label{eqn:rho_out}
\end{align}

From Eq.~(\ref{eqn:rho_out}) we can see how the two techniques used in Sec.~\ref{sec:ancilla_tricks} work.
The biased ancilla of Eq.~(\ref{eqn:ancilla_bias}) can be written as
\begin{equation}
\rho_{\Phi} (x)
\propto
\sum^{\{0,1\}}_{K,K'}
x'_K x'^*_{K'}
\vert \underline{\tilde{K}} \tilde{K} \rangle
\langle \underline{\tilde{K'}} \tilde{K'} \vert,
\label{eqn:ancilla_bias_app}
\end{equation}
where $x'_0 = 1$ and $x'_1 = x$, and it is clear that $x'_K$ remains associated with $\vert \tilde{K} \rangle$ in our final state even after teleportation.

The basis of $\vert \tilde{K} \rangle$ in Eq.~(\ref{eqn:rho_out}) is also evidently arbitrary, as teleportation does not depend in any way on the properties of the ancilla's second mode, and requires only that the first mode be prepared appropriately.
This is similar to the arbitrary order of the third rail seen in the correction scheme proposed by Ref.~\cite{grimsmo2020quantum} using two-mode rotations for the class of RSB codes, of which cat codes are a member.
It may initially appear useful to switch between $R_Z$ and $R_X$ cats to gain access to both Gaussian $Z$ and $X$ operations, respectively, but even in the unitary regime each teleportation risks a full set of Pauli errors that could nullify the desired operation.
However, a teleportation into the GKP code basis, for example, would enable all-optical correction of the $X$ and $Z$ errors via displacements.

\subsection{The generalised syndrome}
\label{app:syndrome}

Here we show the origin of the expression for the syndrome, using an approximation of the channel that considers only pure state cat qubits as inputs and the first $L$ orders of the loss channel.
Otherwise the mechanics are the same as in the prior section, so the derivation will not be exhaustive.
Beginning with the qubit,
\begin{equation}
\vert \psi_{\text{in}} \rangle
=
\frac{q_0 \vert \tilde{0} \rangle + q_1 \vert \tilde{1} \rangle}{\sqrt{2}},
\end{equation}
we can write
\begin{eqnarray}
\vert \psi_{\text{out}} \rangle
\approx
\sum^{\{0,1\}}_{J,K}
\sum^{2L+1}_{j,k=0}
&&
q_J
\,
e^{i j l \pi / (L+1)}
\,
\frac{\bigl( \sqrt{\Gamma} \alpha \bigr)^l}{\sqrt{l!}}
\,
\frac{(-1)^{jJ+kK}}{\sqrt{2 \mathcal{N}_J \underline{\mathcal{N}}_K}}
\,
\frac{e^{(\Gamma-2)\vert \alpha \vert^2 / 2}}{\sqrt{n!m!}}
\,
\left[
\sqrt{\frac{1 - \Gamma}{2}} \alpha
\right]^{n+m}
\nonumber
\\
&&\cdot
\left[ 
e^{i j \pi / (L+1)}
+
e^{i k \pi / (L+1)}
\right]^n
\,
\left[
e^{i k \pi / (L+1)}
-
e^{i j \pi / (L+1)}
\right]^m
\,
\vert \tilde{K} \rangle,
\label{eqn:ket_out}
\end{eqnarray}
where
\begin{equation}
l \equiv L (n+m) \bmod (L+1),
\label{eqn:loss_rule}
\end{equation}
and so the association of $q_J$ with $\vert \tilde{K} \rangle$ allows us to re-write Eq.~(\ref{eqn:ket_out}) as a $2 \times 2$ transformation,
\begin{eqnarray}
S_{KJ} (n,m)
=
\sum^{2L+1}_{j,k=0}
&&e^{i j l \pi / (L+1)}
\,
\frac{\bigl( \sqrt{\Gamma} \alpha \bigr)^l}{\sqrt{l!}}
\,
\frac{(-1)^{jJ+kK}}{\sqrt{2 \mathcal{N}_J \underline{\mathcal{N}}_K}}
\,
\frac{e^{(\Gamma-2)\vert \alpha \vert^2 / 2}}{\sqrt{n!m!}}
\,
\left[
\sqrt{\frac{1 - \Gamma}{2}} \alpha
\right]^{n+m}
\nonumber
\\
&&\cdot
\left[ 
e^{i j \pi / (L+1)}
+
e^{i k \pi / (L+1)}
\right]^n
\,
\left[
e^{i k \pi / (L+1)}
-
e^{i j \pi / (L+1)}
\right]^m.
\end{eqnarray}

\end{widetext}

\bibliography{refs}%

\end{document}